\documentclass{elsart}

\usepackage{amssymb}
\usepackage{graphicx}

\begin{document}

\begin{frontmatter}

\title{Epidemics and chaotic synchronization in recombining monogamous populations}

\author{Federico Vazquez }
\address{Max--Planck--Institut f\"ur Physik Komplexer Systeme \\
N\"othnitzer Str. 38, D--01187 Dresden, Germany\\
and Instituto de F\'{\i}sica Interdisciplinar y Sistemas Complejos
(CSIC-UIB) \\ E-07122 Palma de Mallorca, Spain}

\author{Dami\'an H. Zanette }
\address{Consejo Nacional de Investigaciones
Cient\'{\i}ficas y T\'ecnicas \\ Centro At\'omico Bariloche and
Instituto Balseiro \\ 8400 Bariloche, R\'{\i}o Negro, Argentina}

\begin{abstract}
We analyze the critical transitions (a) to endemic states in an SIS
epidemiological model, and (b) to full synchronization in an
ensemble of coupled chaotic maps, on networks where, at any given
time, each node is connected to just one neighbour. In these
``monogamous'' populations, the lack of connectivity in the
instantaneous interaction pattern --that would prevent both the
propagation of an infection and the collective entrainment into
synchronization-- is compensated by occasional random reconnections
which recombine interacting couples by exchanging their partners.
The transitions to endemic states and to synchronization are
recovered if the recombination rate is sufficiently large, thus
giving rise to a bifurcation as this rate varies. We study this new
critical phenomenon both analytically and numerically.
\end{abstract}


\end{frontmatter}

\section{Introduction}

The spontaneous emergence of different kinds of collective behaviour
is the most paradigmatic phenomenon in natural and artificial
systems formed by large ensembles of interacting dynamical elements.
The interplay of individual dynamics and interactions entrains
elements into coherent macroscopic evolution, which typically
manifests itself in the form of spatial and/or temporal structures.
Pattern formation and synchronization are widespread examples
\cite{col1,col2}.

Self--organization into collective evolution requires that the
information on the state of any single element be able to spread all
over the system, eventually reaching any other element in the
ensemble. A crucial ingredient that controls such mutual influence
of any pair of elements is the interaction pattern of the ensemble.
In the case of binary interactions, this pattern is conveniently
represented as a network, whose links join pairs of elements which
interact with each other \cite{netw}. Coherent evolution of the
whole ensemble is possible only if the interaction network is not
disconnected.

It has recently been shown, in the context of epidemiological
models, that disconnection of the interaction network can however be
compensated to some extent if the structure of the network itself
varies with time, in such a way that different parts of the ensemble
are not continuously but occasionally interconnected
\cite{bouz0,bouz1,bouz2,epjb}. Our aim in this paper is to analyze
in depth the extreme case where, at any given time, each element is
connected by the interaction network to only one neighbour.
Occasional random reconnections can however make neighbour to be
exchanged. As explained in detail in Section \ref{SIS} this scenario
is motivated by the study of a sexually transmitted infection in a
monogamous population, where each individual has just one sexual
partner at a time, but partners can change. Specifically, we focus
on two critical phenomena that occur, upon variation of suitable
control parameters, in a large, connected ensemble of interacting
elements, but which are suppressed if the interaction network only
allows for monogamous static couples. We analyze the reappearance of
each critical phenomenon when random reconnections are allowed to
happen.

In Section \ref{SIS}, we consider the critical transition to endemic
states in an epidemiological model. In Section \ref{Synchr}, we
study the transition to full synchronization in an ensemble of
coupled chaotic maps. Both cases are, to a large extent,
analytically tractable, so that exact results are obtained for the
occurrence of these critical phenomena in time--varying, highly
disconnected networks.

\section{Endemic persistence of SIS epidemics} \label{SIS}

The SIS epidemiological model describes spreading of an infection by
contagion between the members of a population. At any given time,
each agent can be either susceptible (S) or infectious (I). An
S--agent becomes infectious by contact with an I--agent, with
probability $\alpha$ per time unit. An I--agent, in turn,
spontaneously becomes susceptible with probability $\gamma$ per time
unit. Contacts between agents are represented by the links of a
network. For a static random network --provided that there are no
disconnected components, i.e. no portions of the population are
separated from the rest-- the evolution of the fraction $n_{\rm I}$
of I--agents in a large population is well described by the mean
field equation \cite{sis}
\begin{equation}
\dot n_{\rm I} = \beta n_{\rm S}n_{\rm I}- \gamma n_{\rm I},
\end{equation}
where $n_{\rm S}=1-n_{\rm I}$ is the fraction of S--agents, and
$\beta=k \alpha$ is the infectivity, with $k$ the average number of
links per agent in the contact network.

As a function of the infectivity $\beta$, the long--time asymptotic
fraction of I--agents exhibits two distinct regimes. For
$\beta<\gamma$, $n_{\rm I}$ vanishes for long times, and the
infection disappears from the population. For $\beta > \gamma$, on
the other hand, an endemic state with a non--vanishing infection
level is reached. The asymptotic fraction of infected agents in this
state is
\begin{equation} \label{nIs}
n_{\rm I} = 1- \frac{\gamma}{\beta}.
\end{equation}
The transition at the critical point where the infectivity $\beta$
equals the recovery probability $\gamma$ occurs through a
transcritical bifurcation \cite{transcr}.

Let us now assume that we are dealing with a sexually transmitted
infection, where contagion can only occur between sexual partners.
Suppose also that the population is monogamous so that, at any time,
each agent has just one partner. For simplicity, we assume that all
agents have partners. In this situation, the network of contacts is
highly disconnected: for a population comprising  $N$ agents, the
network consists of $N/2$ isolated links defining agent couples. If
at least one agent is initially infectious within a given couple,
the infection can persist for some time as a result of repeated
contagion between the two partners. However, there is a finite
probability that in any finite time interval both agents become
susceptible. Assuming that no contacts are allowed with other
members of the population, the two recovered partners will never
become infectious again. If couples last forever, consequently, the
fate of the sexually transmitted infection is to disappear from the
population. Due to its lack of connectivity, the contact network is
unable to sustain a long--time endemic state.

If, on the other hand, sexual partners are allowed to change from
time to time, even if at any instant the population is still
monogamous, the infection may spread over the population and,
eventually, reach an endemic level. Specifically, let us consider
that each couple ($i,j$) can exchange partners with another randomly
selected couple ($h,l$) at a given rate, creating new couples
($i,h$) and ($j,l$). Would it be possible that, if these
recombination events are frequent enough, an endemic state is
established in the monogamous population? To investigate this
question we adopt two different approaches, which can to a large
extent be dealt with analytically. In the first one \cite{epjb}, we
compute the fraction of I--agents from the contribution of couples
formed at different times in the past. In the second approach, we
study the evolution of the number of couples containing two, one, or
no infectious partners.

\subsection{Epidemiological dynamics within couples} \label{2.1}

Since the moment when two given agents become joined in a couple,
the probabilities of their being either susceptible or infectious
evolve independently of the rest of the population. Let $p_0(t)$,
$p_1(t)$, and $p_2(t)$ be the probabilities that, at time $t$, the
couple under study comprises zero, one, and two I--agents,
respectively. These probabilities satisfy the equations
\begin{equation} \label{p0p1p2}
\begin{array}{ll}
\dot p_0 &= \gamma p_1, \\
\dot p_1 &= 2\gamma p_2-(\gamma+\beta) p_1 , \\
\dot p_2 & = \beta p_1 -2\gamma p_2 .
\end{array}
\end{equation}
At all times, $p_0+p_1+p_2=1$, so that  the analysis can be
restricted to the system formed by the two last lines of the above
equations. As expected, the only fixed point of this linear system,
$(p_1,p_2)=(0,0)$, corresponds to the disappearance of the
infection. The eigenvalues around the fixed point are
\begin{equation}
\mu_\pm = -\frac{1}{2} (\beta+3 \gamma) \pm  \frac{1}{2}
\sqrt{\beta^2+6 \beta \gamma+\gamma^2}.
\end{equation}
Note that $\mu_-<\mu_+<0$. The inverse modulus of $\mu_+$ gives the
typical duration time of the infection within a couple, $T_{\rm I}$.
For $\gamma \gg \beta$ and  $\gamma \ll \beta$ we have,
respectively, $T_{\rm I} \sim 1/\gamma$ and $T_{\rm I} \sim
\beta/\gamma^2$. The general solution for $p_1$ and $p_2$ can be
written as $(p_1,p_2) = A_+ {\bf e}_+ \exp(\mu_+ t) + A_- {\bf e}_-
\exp(\mu_- t) $, with
\begin{equation}
{\bf e}_\pm = \left( 1, \frac{\beta}{2\gamma + \mu_\pm} \right),
\end{equation}
and where the coefficients $A_\pm$ are determined by the initial
values of $p_1$ and $p_2$.

When a couple is formed, the initial values of $p_1$ and $p_2$ can
be evaluated, on the basis of mean field--like arguments, from the
fraction $n_{\rm I}$ of I--agents at that time: $p_1 = 2n_{\rm I}
(1-n_{\rm I})$ and $p_2=n_{\rm I}^2$. Reversing the same arguments,
the linear combination $p_1+2p_2$ gives the expected fraction of
I--agents in the same couple. With these elements, we can now write
down an equation for the evolution of $n_{\rm I} (t)$ taking into
account couple recombination events. The fraction of I--agents at
the present time $t$ is given by the contributions of all the
present couples, formed at previous times $t-s$ ($0\le s\le t$). The
initial probabilities within each couple are determined by the
fraction of $I$--agents at the time of formation, $n_{\rm I}(t-s)$,
and the present probabilities are given by the solution to Eqs.
(\ref{p0p1p2}) discussed above. Summing up all the contributions, we
get
\begin{eqnarray} \label{nI}
n_{\rm I} (t) = &\int_0^t \Pi(t,s) \left[ \frac{1}{2}\left( 1+
\frac{2\beta}{2\gamma+\mu_-}\right) A_-[n_{\rm I}(t-s)] \exp(\mu_-
s) \right. \nonumber
\\ &\left.  + \frac{1}{2}\left( 1+ \frac{2\beta}{2\gamma+\mu_+}\right) A_+
[n_{\rm I}(t-s)] \exp(\mu_+ s)\right] ds .
\end{eqnarray}
The coefficients $A_\pm (n_{\rm I})$ are obtained from the
evaluation of the initial probabilities in a couple in terms of the
fraction of I--agents at that time:
\begin{equation}
\left(2n_{\rm I} (1-n_{\rm I}),  n_{\rm I}^2 \right) = A_-(n_{\rm I}
) {\bf e}_- + A_+(n_{\rm I} ) {\bf e}_+.
\end{equation}

In Eq. (\ref{nI}), the contribution to $n_{\rm I} (t)$ of the
couples formed at $t-s$ is weighted by $\Pi(t,s)$, the probability
that a couple present at time $t$ has lasted for an interval of
length $s$. Since couple recombination occurs at random, this
probability distribution is a Poissonian function of $s$,
specifically,
\begin{equation}
\Pi(t,s) = \frac{r \exp(-rs)}{1-\exp(-rt)},
\end{equation}
for $0\le s\le t$. Here, $r$ is the recombination probability per
couple per time unit.

The rather involved form of our equation for $n_{I} (t)$, Eq.
(\ref{nI}), is drastically simplified in the long--time limit. We
find that for small values of the recombination rate $r$, the
asymptotic fraction of I--agents vanishes --just as when
recombination is absent and the infection, confined within couples,
eventually disappears. There is however a critical value of the
recombination rate above which an endemic state is reached. The
stationary fraction of I--agents is
\begin{equation} \label{nIa}
n_{\rm I}= 1-\frac{\gamma}{\beta} \left( 1+\frac{2\gamma}{r}\right).
\end{equation}
The appearance of the endemic state occurs here through a
transcritical bifurcation, like when the infectivity $\beta$ is
varied in the standard SIS model.  The critical value of the
recombination rate is
\begin{equation} \label{critic}
r_c = \frac{2\gamma}{\beta/\gamma-1}.
\end{equation}
Note that Eq. (\ref{nIa}) reduces to Eq. (\ref{nIs}) in the limit
$r\to \infty$. For a given value of the infectivity $\beta$, thus,
infinitely frequent recombination in a monogamous contact pattern
sustains a stationary infection level equivalent to that of a
population with a static, not disconnected pattern.

\begin{figure} \centering
\resizebox{\columnwidth}{!}{\includegraphics{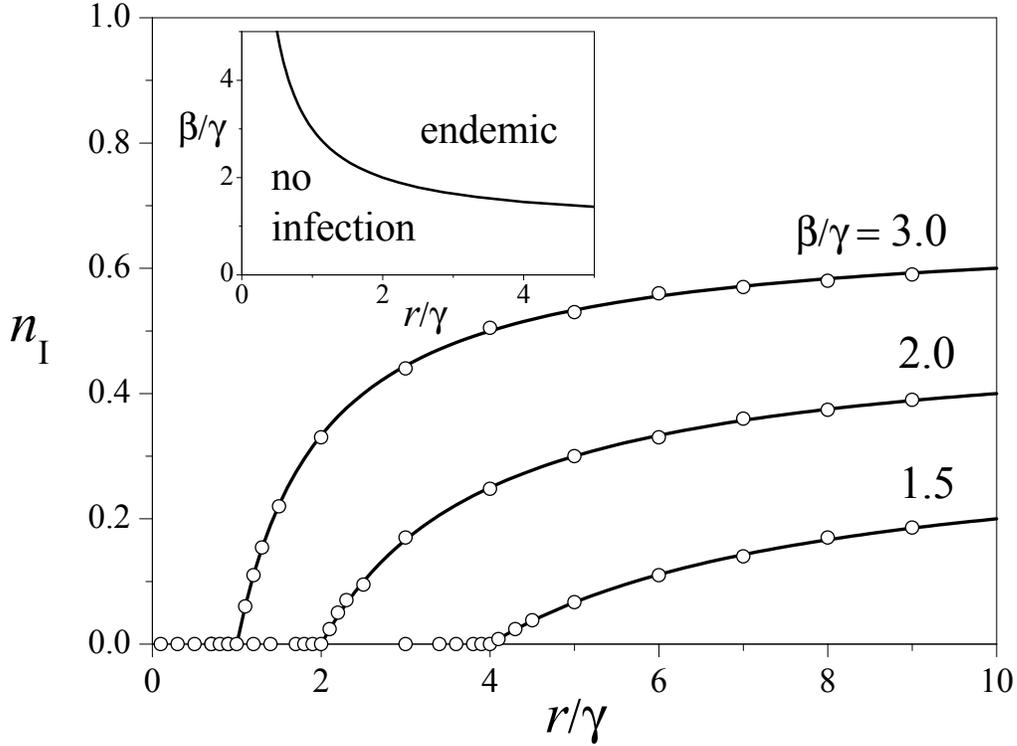}} \caption{
Long--time asymptotic fraction of infectious agents $n_{\rm I}$, as
a function of the recombination rate $r$ (normalized by the recovery
probability $\gamma$), for three values of the normalized
infectivity $\beta/\gamma$. Curves and dots correspond,
respectively, to analytical and numerical results. The inset shows,
in the parameter plane ($r/\gamma, \beta/\gamma$), the boundary
between the regimes of endemic infection and infection
disappearance.} \label{fig1}
\end{figure}

To validate the arguments used to derive Eqs. (\ref{nI}) and, in
particular, the asymptotic result (\ref{nIa}), we have performed
numerical simulations of the epidemiological model in a recombining
monogamous population of $N=10^3$ agents. Since two couples are
involved in each recombination event, the probability per time unit
of each event in our numerical simulations equals $r/2$. Figure
\ref{fig1} compares analytical and numerical results, which turn out
to show excellent agreement. The inset shows the boundary between
the regions of parameter space where, for asymptotically long times,
the infection persists or disappears, as given by Eq.
(\ref{critic}).

\subsection{Evolution of the number of couples} \label{2.2}

Equivalent asymptotic results are obtained from mean field equations
for the number of couples formed by agents with different
epidemiological states. Note that, in our problem, the dynamics of
the number of couples of different kinds corresponds to the
evolution of links in the contact network. Let $m_{\rm SS}$, $m_{\rm
IS}$, and $m_{\rm II}$ be the fraction of couples formed,
respectively, by two S--agents, one I--agent and one S--agent, and
two I--agents. Since at all times $m_{\rm SS}+m_{\rm IS}+ m_{\rm
II}=1$, it is enough to consider the evolution of only two of them,
for instance, $m_{\rm SS}$ and $m_{\rm II}$. The mean field
equations read
\begin{equation} \label{couples}
\begin{array}{ll}
\dot m_{\rm SS} & = \frac{r}{4} m^2_{\rm IS} - r m_{\rm SS} m_{\rm
II}+ \gamma m_{\rm IS} , \\
\dot m_{\rm II} & = \frac{r}{4} m^2_{\rm IS} - r m_{\rm SS} m_{\rm
II}+ \beta m_{\rm IS}- 2\gamma m_{\rm II},
\end{array}
\end{equation}
with  $m_{\rm IS}=1- m_{\rm SS}- m_{\rm II}$. The first two terms in
the right--hand side of both equations stand for the only
recombination events that change the number of couples of each kind,
namely, (SS,II) $\leftrightarrow$ (IS,IS). The remaining terms
correspond to the epidemiological events of contagion and recovery.

Equations (\ref{couples}) have two stationary solutions. One of them
corresponds to the disappearance of the infection:
\begin{equation}
m_{\rm SS} = 1, \ \ \ \ \ m_{\rm II} = 0.
\end{equation}
The other fixed point,
\begin{equation}
m_{\rm SS} = \frac{\gamma^2}{\beta^2 r} (2\beta+2\gamma+r), \ \ \ \
\ m_{\rm II} = \frac{\beta-\gamma}{\beta^2 r} [(\beta-\gamma)
r-2\gamma^2 ],
\end{equation}
corresponds to the endemic state. In agreement with the results of
Section \ref{2.1}, the two stationary solutions interchange
stability when the recombination rate attains the critical value
given by Eq. (\ref{critic}). The infection level in the endemic
state can be evaluated from the number of couples as $n_{\rm I}
\equiv m_{\rm II} + m_{\rm IS}/2 = (1+m_{\rm II}-m_{\rm SS})/2$ and
coincides with the fraction of I--agents of Eq. (\ref{nIa}).

\section{Synchronization of coupled chaotic maps} \label{Synchr}

The reappearance of the transition to endemic states in the
monogamous SIS epidemiological model upon couple recombination
events, opens the question of whether recombination may compensate
disconnection for the occurrence of similar critical phenomena in
other ensembles of interacting dynamical elements. In this section,
we explore this question for the synchronization transition of
coupled chaotic maps.

Consider an ensemble of $N$ identical chaotic maps whose individual
states $x_i$ ($i=1,\dots, N$) evolve, in the absence of coupling,
according to $x_i(t+1)= f[x_i(t)]$. Let $\lambda>1$ be the
corresponding Lyapunov coefficient. On the average, the distance
between two neighboring orbits of a map thus grows by a factor
$\lambda$ at each iteration. Global (all-to-all) coupling between
maps is introduced following the standard linear scheme
\cite{Kaneko,Kanekoec}
\begin{equation} \label{couplN}
x_i(t+1)= (1-\varepsilon)  f[x_i(t)] + \frac{\varepsilon}{N}
\sum_{j=1}^N f[x_j (t)],
\end{equation}
where $\varepsilon \in [0,1]$ is the coupling intensity. It is well
known that, if coupling is strong enough, the ensemble undergoes a
transition to full synchronization. Specifically, if
\begin{equation}
\varepsilon > \varepsilon_c = 1-\frac{1}{\lambda},
\end{equation}
coupling is able to overcome the exponential separation of chaotic
orbits, and all the maps converge asymptotically to exactly the same
trajectory: $x_i(t)=x_j(t)$ for all $i$, $j$ and $t\to \infty$
\cite{Kanekoec}. This synchronized trajectory is still chaotic --in
fact, it reproduces the orbit of a single map-- but the motion of
the maps with respect to each other has been suppressed. Note that
the threshold $\varepsilon_c$ for the synchronization transition
does not depend on the ensemble size $N$.

Consider now a ``monogamous'' coupling pattern where, at any given
time, each map is coupled to just one partner. Following the scheme
of Eq. (\ref{couplN}), if maps $a$ and $b$ form a couple, their
states evolve according to
\begin{equation} \label{coupl2}
\begin{array}{ll}
x_a(t+1) &= (1-\varepsilon)  f[x_a(t)]+ \frac{\varepsilon}{2}
\left\{ f[x_a(t)]+ f[x_b(t)]\right\} ,  \\
x_b(t+1) &= (1-\varepsilon) f[x_b(t)]+ \frac{\varepsilon}{2} \left\{
f[x_a(t)]+ f[x_b(t)]\right\} .
\end{array}
\end{equation}
If $\varepsilon > \varepsilon_c$, the two maps will synchronize,
converging asymptotically to the same chaotic orbit. Because of the
nature of chaotic motion, however, their synchronized trajectory
will differ from the orbits of other couples.

Would it be possible that, if couples of maps recombine exchanging
partners at random at a certain rate, full synchronization all over
the ensemble is recovered? By analogy with the occurrence of endemic
states in SIS epidemics over a recombining monogamous population, we
may expect that synchronization spreads over the ensemble if
recombination is faster enough, entraining increasingly many maps
towards the fully synchronized orbit. However, a difference with the
case of SIS epidemic is that, now, the recombination rate has a
limit: one random recombination per couple per iteration step
produces the maximal possible rearrangement of the coupling pattern.

\subsection{Synchronization dynamics within couples}

To investigate whether full synchronization is possible in the
monogamous coupling pattern under the effects of recombination, we
assume that the ensemble is symmetrically concentrated around a
reference orbit $x_0(t)$ of the map $f(x)$, so that the state of
each map differs from $x_0(t)$ by a small amount:
$x_i(t)=x_0(t)+\delta_i(t)$. If, as time elapses, the concentration
around $x_0(t)$ grows in such a way that $\delta_i(t) \to 0 $ for
all $i$, full synchronization will be achieved. Equations
(\ref{coupl2}) imply that the displacements of two coupled maps from
the reference orbit $x_0(t)$ satisfy the linear equations
\begin{equation} \label{deltaij}
\begin{array}{ll}
\delta_a(t+1)  &= (1-\frac{\varepsilon}{2}) f'[x_0(t)] \delta_a (t)
+ \frac{\varepsilon}{2} f'[x_0(t)] \delta_b(t),
\\ \delta_b(t+1)  &= (1-\frac{\varepsilon}{2}) f'[x_0(t)] \delta_b (t) +
\frac{\varepsilon}{2} f'[x_0(t)] \delta_a (t),
\end{array}
\end{equation}
where $f'(x)$ is the derivative of $f(x)$.

Let us first study the simpler, extreme case where all couples
recombine at each iteration step. Since new partners are chosen at
random, the displacements $ \delta_a (t)$ and  $\delta_b (t)$ of the
two coupled maps are uncorrelated quantities. In other words, the
average $\zeta (t) = \langle \delta_a (t)  \delta_b (t) \rangle$,
performed over all the couples in the ensemble, vanishes.
Consequently, from Eqs. (\ref{coupl2}), the variance of the
displacement over the ensemble, $\sigma^2 (t) = \langle \delta_a^2
(t) \rangle$, is governed by
\begin{equation}
\sigma^2 (t+1)  = \left(1-\varepsilon+\frac{\varepsilon^2}{2}
\right) f'[x_0(t)]^2 \sigma^2 (t).
\end{equation}
Taking into account that the long--time geometric mean value of
$f'[x_0(t)]^2$ is
\begin{equation} \label{lim}
\lim_{T \to \infty} \left( \prod_{t=t_0}^{t_0+T} f'[x_0(t)]^2
 \right)^{1/T} = \lambda^2,
\end{equation}
the variance $\sigma^2$ will asymptotically vanish if
$(1-\varepsilon+\varepsilon^2/2) \lambda^2 <1$ or, equivalently, if
\begin{equation} \label{epsilon1}
\varepsilon >  \varepsilon_1 = 1-\frac{\sqrt{2-\lambda^2}}{\lambda}.
\end{equation}
Thus, if the coupling intensity $\varepsilon$ is larger than the
critical value $\varepsilon_1$, full synchronization is stable. Note
that $\varepsilon_1 > \varepsilon_c$ for any $\lambda \gtrsim 1$, so
that full synchronization for maximal recombination rate requires
stronger couplings than when the ensemble is globally coupled. Also,
there is a critical limit $\lambda_1 =\sqrt{2}\approx 1.41$ for the
Lyapunov coefficient such that, if $\lambda > \lambda_1$, full
synchronization is impossible even under the maximal recombination
rate.

When recombination does not occur for all couples at all times, the
joint evolution of two coupled maps introduces correlations between
their displacements from the reference orbit. In this case, Eqs.
(\ref{coupl2}) imply that $\zeta (t) = \langle \delta_a (t) \delta_b
(t) \rangle$ and $\sigma^2 (t) = \langle \delta_a^2 (t) \rangle$ are
governed by
\begin{equation}\label{sigmazeta}
\begin{array}{ll}
\zeta(t+1)  &= (1-\frac{\varepsilon}{2}) \varepsilon  f'[x_0(t)]^2
\sigma^2 (t)+
(1-\varepsilon+\frac{\varepsilon^2}{2}) f'[x_0(t)]^2 \zeta(t), \\
\sigma^2 (t+1)  &= (1-\varepsilon+\frac{\varepsilon^2}{2})
f'[x_0(t)]^2 \sigma^2 (t)+ (1-\frac{\varepsilon}{2}) \varepsilon
f'[x_0(t)]^2 \zeta(t).
\end{array}
\end{equation}
This linear system can be solved exactly. For the couples formed at
a certain time $t_0$, we have $\zeta (t_0)=0$ and $\sigma^2(t_0)
=\sigma_0^2$. With such initial conditions, and taking into account
Eq. (\ref{lim}), the solution to Eqs. (\ref{sigmazeta}) for
$\sigma^2$ reads
\begin{equation} \label{sigma}
\sigma^2(t) = \frac{1}{2} \lambda^{2(t-t_0)}
\left[1+(1-\varepsilon)^{2 (t-t_0)} \right]  \sigma_0^2,
\end{equation}
for  $t\gg t_0$. This quantity is the variance of the displacement
from $x_0(t)$ of those maps whose present couples have formed at
time $t_0$ and lasted until the present time $t$.

Suppose now that, at a certain time $\tau$, the ensemble around
$x_0(\tau)$ has variance $\sigma_\tau^2$. To evaluate the variance
$\sigma_{\tau+1}^2$ at the next time step, we think of the ensemble
as made up of subensembles consisting of the maps whose present
couples have formed at times $\tau$, $\tau-1$, $\tau-2$, and so on.
The fraction of maps in the subensemble corresponding to couples
formed at time $\tau-n$ ($n=0,1,2,\dots$) is $q_n = r (1-r)^n$,
where $r$ is the probability that any couple forms at any given time
step ($0 \le r\le 1$). Using Eq. (\ref{sigma}) with $t \equiv \tau $
and $t_0 \equiv \tau - n$, we find that the variance of this
subensemble, which we denote $\sigma_n^2$, changes from
$\sigma_n^2(\tau) = \sigma_\tau^2$ to
\begin{equation}
\sigma_n^2 (\tau+1) = \lambda^2
\frac{1+(1-\varepsilon)^{2(n+1)}}{1+(1-\varepsilon)^{2n}}
\sigma_\tau^2 .
\end{equation}
Assuming that the ensemble has been evolving since an asymptotically
long time ago, its variance at time $\tau +1$ is given by
\begin{equation}
\sigma_{\tau+1}^2 = \sum_{n=0}^\infty q_n \sigma_n^2 (\tau+1)=
r\lambda^2 \sum_{n=0}^\infty (1-r)^n
\frac{1+(1-\varepsilon)^{2(n+1)}}{1+(1-\varepsilon)^{2n}}
\sigma_\tau^2 .
\end{equation}
The synchronization threshold  is therefore given by the condition
\begin{equation} \label{synchr}
r\lambda^2 \sum_{n=0}^\infty (1-r)^n
\frac{1+(1-\varepsilon_r)^{2(n+1)}}{1+(1-\varepsilon_r)^{2n}} =1,
\end{equation}
where $\varepsilon_r$ is the critical coupling intensity.

Note that, to obtain this result, we have used Eq. (\ref{sigma}) for
$t-t_0=n=0,1,2,\dots$ while, strictly, it holds for $t \gg t_0$
only. Consequently, the threshold condition (\ref{synchr}) is just
an approximation whose validity must be ascertained for each
specific choice of the map $f(x)$. Below, we present numerical
results for a case where Eq. (\ref{sigma}) holds at any time.

Whereas, in general, the summation in Eq. (\ref{synchr}) cannot  be
exactly performed, two special cases are readily obtained. The first
one corresponds to the maximal recombination rate, $r=1$, for which
we reobtain the critical coupling intensity $\varepsilon_1$ of Eq.
(\ref{epsilon1}). The second special case gives the recombination
rate for which the threshold coupling intensity is maximal,
$\varepsilon_r=1$, namely,
\begin{equation}
r_{\min}=2\left( 1-\frac{1}{\lambda^2} \right).
\end{equation}
For recombination rates below $r_{\min}$, synchronization is not
possible even for the strongest coupling.

\subsection{Numerical simulations}

To test the above results we have performed numerical simulations of
recombining monogamous ensembles of $N=10^3$ chaotic maps for the
case of the tent map \cite{transcr}:
\begin{equation}
f(x) = \left\{
\begin{array}{ll}
px & \mbox{for $0\le x \le \frac{1}{2}$}, \\
p(1-x) & \mbox{for $\frac{1}{2}\le x \le 1$},
\end{array}
\right.
\end{equation}
with $ x \in [0,1]$ and $p \in [0,2]$. The Lyapunov coefficient of
the tent map is $\lambda = p$, which implies that the dynamics is
chaotic for $1<p\le 2$. Moreover, $f'(x)^2 = p^2 = \lambda^2 $ for
all $x$, so that in Eq. (\ref{lim}) the limit of $T\to \infty$ can
be dropped, and Eq. (\ref{sigma}) holds for any $t$.

In our simulations, after joint evolution of the coupled maps
following Eq. (\ref{coupl2}), each one of the $N/2$ couples is
allowed to recombine with probability $\rho$, exchanging partners
with another randomly selected couple. Since two couples are
involved at each recombination event, the resulting recombination
probability per couple per time step is larger than $\rho$, and
reads
\begin{equation} \label{rho}
r=1-(1-\rho) \exp (-\rho).
\end{equation}
Fully synchronized ensembles are detected by numerically measuring
the variance $\sigma_x^2 (t) = N^{-1} \sum_i [x_i(t)-\bar x(t)]^2$
with respect to the average state $\bar x(t) = N^{-1} \sum_i
x_i(t)$. Realizations where $\sigma_x^2 (t)$ falls to the level of
numerical round-off errors are identified with full synchronization.

\begin{figure} \centering
\resizebox{\columnwidth}{!}{\includegraphics{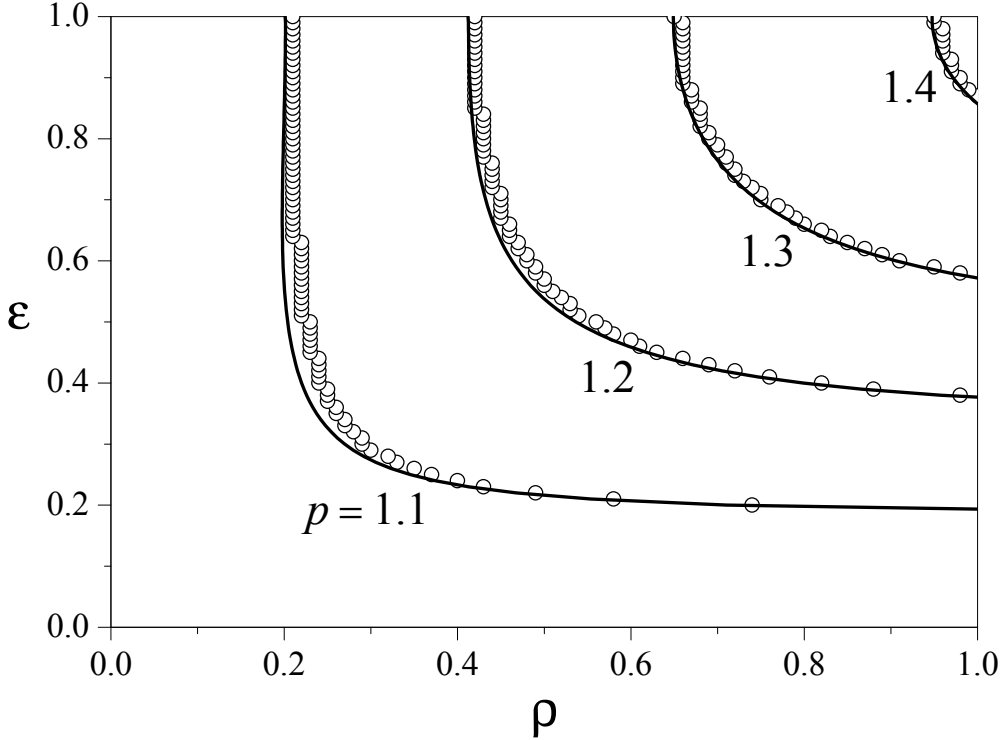}}
\caption{Full-synchronization threshold for a recombining monogamous
ensemble of  tent maps, in the parameter plane $(\rho,\varepsilon)$,
for four values of the map parameter $p$. Full synchronization is
stable above and to the right of each boundary. Dots and curves
stand, respectively, for numerical results on ensembles of $10^3$
maps and the analytical result of Eq. (\ref{synchr}). The numerical
parameter $\rho$ is related to the recombination rate $r$ through
Eq. (\ref{rho}).} \label{fig2}
\end{figure}

Figure \ref{fig2} shows the boundary of full synchronization in the
parameter space $(\rho,\varepsilon)$ for four values of the tent map
slope $p$. For each value of $\rho$, full synchronization is stable
for coupling intensities $\varepsilon$ above the boundary. Dots
stand for the numerical determination of the synchronization
threshold and curves correspond to the analytical prediction, Eq.
(\ref{synchr}). The agreement is very good. The systematic
difference between numerical and analytical results, more visible
for small $p$, is to be ascribed to the numerical discretization of
parameter space in the search for the synchronization threshold.

\begin{figure} \centering
\resizebox{\columnwidth}{!}{\includegraphics{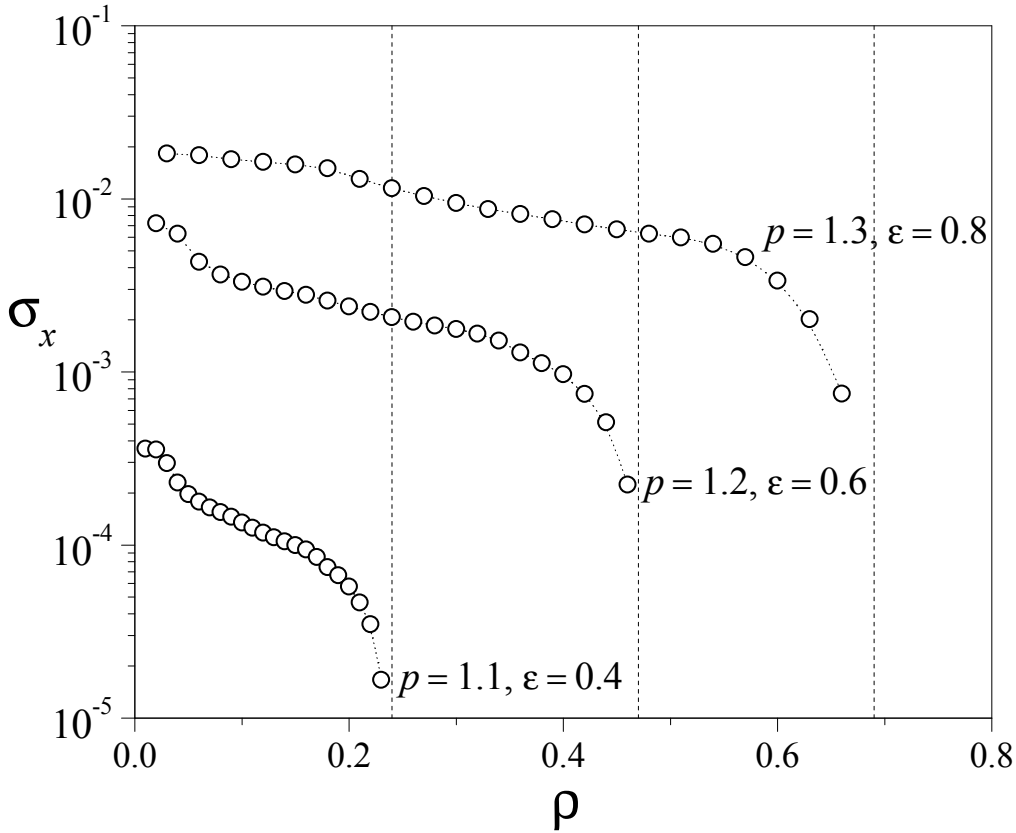}}
\caption{Dispersion of a recombining monogamous ensemble of $N=10^3$
 tent maps as a function of the recombination
parameter $\rho$, for three values of $p$ and $\varepsilon$.
Vertical dashed lines stand for the numerical determination of the
critical value of $\rho$, for each combination of $p$ and
$\varepsilon$. Dotted curves are splines added as a guide to the
eye.} \label{fig3}
\end{figure}

In Fig. \ref{fig3}, we present numerical measurements of the
long-time asymptotic values of the dispersion $\sigma_x$ as a
function of $\rho$, for three combinations of $p$ and $\varepsilon$.
Note that the drop of $\sigma_x$ at the critical recombination rate
is abrupt but continuous.

\subsection{Two-state symbolic dynamics}

A description of the dynamics of the ensemble of coupled chaotic
maps in terms of the evolution of different kinds of couple, along
the lines developed in Section \ref{2.2} for SIS epidemics, is in
principle not possible. In fact, while the agents in the SIS
population admit just two possible individual states --either
susceptible or infectious, which determine three kinds of couple--
the individual state of each chaotic map can span a continuous
interval. However, as we show in this section, a symbolic
representation of each map as a two-state variable --either
synchronized or unsynchronized-- together with a reduced set of
transition rules between the two states, is able to qualitatively
reproduce the collective effects of recombination on the
synchronization transition. This schematic approach has the virtue
of capturing the essential mechanisms of the interplay between
recombination and synchronization, and can also be reinterpreted in
the context of the SIS epidemics model.

Let us thus assume that, at any given time, each chaotic map adopts
one of two states, namely, unsynchronized (U) or synchronized (S).
We stress that synchronization of an individual map $i$ is here
understood as defined with respect to the ensemble, for instance, if
the variable $x_i (t)$ is within a certain small distance from the
average $\bar x(t) = N^{-1} \sum_j x_j(t)$. In the spirit of Eqs.
(\ref{couples}), we propose for the evolution of the fractions of
couples $m_{\rm UU}$, $m_{\rm US}$, and $m_{\rm SS}$ the Ansatz
\begin{equation} \label{coupless}
\begin{array}{ll}

\dot m_{\rm UU} & = \frac{r}{4} m^2_{\rm US} - r m_{\rm SS} m_{\rm
UU}+ \gamma m_{\rm US}- \eta(r) m_{\rm UU},
\\
\dot m_{\rm SS} & = \frac{r}{4} m^2_{\rm US} - r m_{\rm SS} m_{\rm
UU}+ \beta (r) m_{\rm US} + \eta(r) m_{\rm UU},

\end{array}
\end{equation}
with $m_{\rm US}=1-m_{\rm SS}-m_{\rm UU}$, and where $\beta (r)$,
$\gamma$, and $\eta (r)$ are non--negative quantities. The first two
terms in the right--hand side of both equations describe the effect
of recombination at rate $r$ per couple. They have exactly the same
origin as in Eqs. (\ref{couples}). The remaining terms, described in
detail in the following, stand for the transitions induced by the
joint evolution of coupled maps. Note that we are assuming that all
these transitions occur always toward states where the two maps of a
couple are either synchronized or unsynchronized. Also, we suppose
that the two maps of an SS--couple, as long as their link lasts,
cannot become unsynchronized from the ensemble.

In the first place, as time elapses, a US--couple can become either
an SS--couple or a UU--couple. This describes the tendency towards
synchronization within each couple. The rate for the transition  US
$\to$ SS, which we have called $\beta$ in Eqs. (\ref{coupless}),
cannot however be a constant. If it were a constant, in fact, all
US--couples would become SS--couples for sufficiently long times,
thus synchronizing with the ensemble even in the absence of
recombination. This is not the case, though: when recombination is
infrequent, maps forming a long-lasting couple synchronize to each
other, but are generally not synchronized to the ensemble. We
represent this effect phenomenologically, by ascribing to $\beta$ a
dependence on $r$. The function $\beta (r)$ increases as $r$ grows,
starting from $\beta (0) = 0$.

The transition  US $\to$ UU, on the other hand, implies the
desynchronization of a map with respect to the ensemble by
interaction with its already unsynchronized partner. This event does
not require recombination: it is rather the generally expected
outcome within a US couple. We thus assume that its rate, $\gamma$,
is constant.

Finally, Eqs. (\ref{coupless}) include, in the last terms of their
right--hand side, the transition  UU $\to$ SS. Because of the same
reasons as in the case of the transition  US $\to$ SS, which also
involves the synchronization with the ensemble of previously
unsynchronized maps, we expect that the corresponding rate, $\eta
(r)$, is a growing function of the recombination rate $r$. Possibly,
however, $\eta (r)$ is much smaller that $\gamma (r)$, because
becoming an SS--couple should occur less frequently for a UU--couple
than for a US--couple.

Equations (\ref{coupless}) have two fixed points. One of them,
$m_{\rm SS} =1$, $m_{\rm UU} =0$, corresponds to full
synchronization of the ensemble of chaotic maps. For the other fixed
point, both $m_{\rm SS} $ and $m_{\rm UU}$ are different from zero,
thus corresponding to a state of partial synchronization. There is a
broad choice of functional forms for $\beta (r)$ and $\eta (r)$ such
that these stationary solutions have the expected behaviour
--namely, that full synchronization is unstable for small $r$ and
becomes stable above a critical value of the recombination rate,
while the other fixed point exhibits opposite stability properties.
For the sake of concreteness, let us assume the linear dependences
$\beta (r) = \beta_0 r$ and $\eta (r) =\eta_0 r$. Analysis of the
eigenvalues of Eqs. (\ref{coupless}) around the fixed points shows
that, irrespectively of the values of $\beta_0$ and $\gamma$, it is
enough that $\eta_0 < 1$ for the occurrence of a transition to full
synchronization as $r$ grows. The critical value of the
recombination rate is
\begin{equation} \label{rcs}
r_c = \frac{\gamma}{\beta_0} \frac{1-\eta_0}{1+\eta_0}.
\end{equation}

\begin{figure} \centering
\resizebox{\columnwidth}{!}{\includegraphics{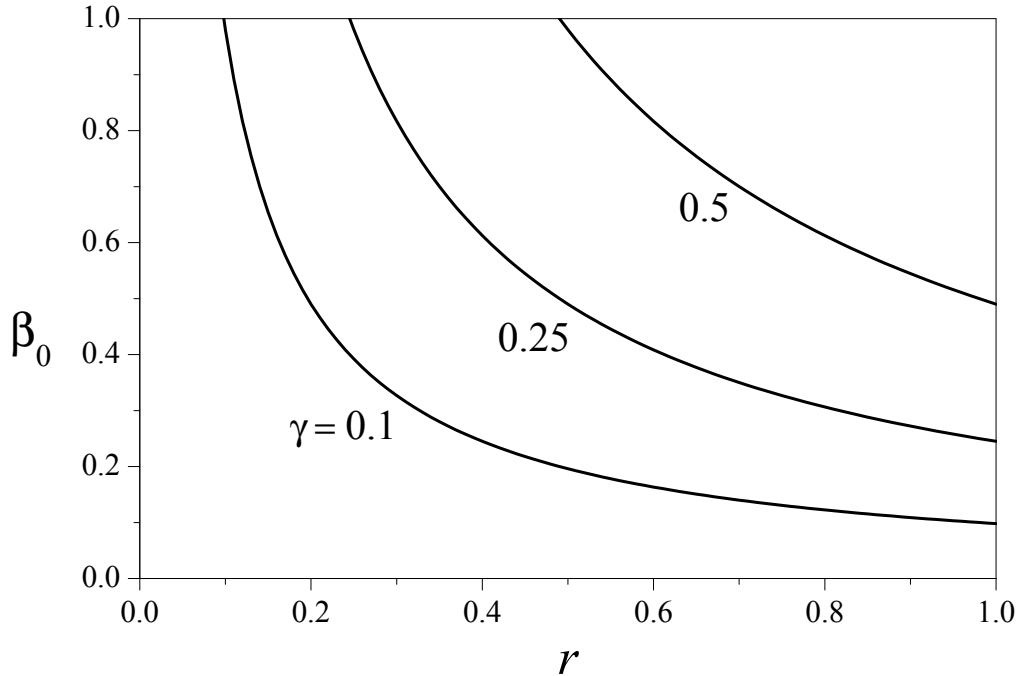}}
\caption{Synchronization threshold in the parameter plane
($r,\beta_0$) for the two--state model of coupled chaotic maps,
corresponding to three values of the rate $\gamma$ and
$\eta_0=0.01$. Full synchronization is stable in the region of large
$r$ and $\beta_0$. Compare this plot with Fig. \ref{fig2}.}
\label{fig4}
\end{figure}

Let us now attempt a semi--quantitative comparison of this result
with our results for the synchronization transition in recombining
monogamous ensembles of chaotic maps, for instance, those depicted
in Fig. \ref{fig2} for tent maps. Since $\gamma$ is the rate for the
transition  US $\to$ UU --where, by interaction within its couple, a
map becomes desynchronized from the ensemble-- it may be identified
with a measure of the rate at which chaotic orbits diverge from each
other, i.e. of the Lyapunov coefficient $\lambda$. The factor
$\beta_0$, in turn, weights the rate of the transition US $\to$ SS,
where a map becomes synchronized to the ensemble by interaction with
its already synchronized partner. Thus, $\beta_0$ is related to the
strength of the interaction between coupled maps, i.e. to the
coupling intensity $\varepsilon$. The factor $\eta_0$ plays a
similar role but, as mentioned above, its effect should be
quantitatively less important than that of $\beta_0$. Figure
\ref{fig4} is a plot of the synchronization threshold as given by
Eq. (\ref{rcs}). To stress the correspondence with Fig. \ref{fig2},
we have plotted $\beta_0$ (a measure of the coupling intensity
$\varepsilon$) as a function of the recombination rate $r$ (directly
related, through Eq. (\ref{rho}), to the  parameter $\rho$) for
three values of $\gamma$ (a measure of the Lyapunov coefficient,
$\lambda=p$ for the tent map), and fixed $\eta_0$. The
semi--quantitative analogy with our numerical results for ensembles
of tent maps, shown in Fig. \ref{fig2},  is apparent.

An analogy between this two--state approach to synchronization and
the SIS epidemiological model discussed in Section \ref{SIS} can in
turn be derived from the identification of the unsynchronized state
with the susceptible state on one hand, and of the synchronized
state and the infectious state on the other. With this
identification, the transitions US $\to$ SS and US $\to$ UU
correspond, respectively, to contagion from the infectious partner
and to the spontaneous recovery of an infectious agent. The rates
$\beta$ and $\gamma$ play the same role in both dynamical models,
with the difference that in the SIS model $\beta$ does not depend on
the recombination rate. Another difference is that the two--state
approach to synchronization includes the transition  UU $\to$ SS,
which in the SIS model would correspond to simultaneous infection of
two susceptible partners --a forbidden event. The SIS model, in
turn, allows for the recovery of just one among two infectious
partners, which would stand for the inexistent transition SS $\to$
US. Notwithstanding these differences, we realize that the plot of
Fig. \ref{fig4} is the equivalent in the two--state approach to
synchronization as the endemic threshold depicted in the insert of
Fig. \ref{fig1} --where both the infectivity $\beta$ and the
recombination  $r$ are normalized by the recovery probability
$\gamma$. The two thresholds have qualitatively the same functional
dependence on the respective parameter. In addition, the transition
to full synchronization in our two--state formulation is a
transcritical bifurcation, the same kind of critical phenomenon as
in the SIS model.

\section{Conclusion}

We have here studied two critical phenomena in the collective
behaviour of large ensembles of interacting dynamical elements
--namely, the appearance of endemic states in an SIS epidemiological
model, and the stabilization of full synchronization of identical
chaotic maps-- when the corresponding interaction patterns are
highly disconnected and, concurrently, change with time.
Specifically, we have considered ```monogamous'' interaction
patterns, represented by networks where each site is connected to
only one neighbour at a time, but such that neighbours can be
exchanged at random at a specified rate. While in the absence of
neighbour exchange --or, as we have called it, of recombination--
the occurrence of endemic states and synchronization would be
impossible due to the lack of connectivity in the ensemble,
sufficiently frequent recombination events make it possible that
coherence is established all over the system, thus allowing for
organized collective dynamics.

Recombination of interacting couples in a monogamous pattern
introduces a new dynamical parameter --the recombination rate. The
two critical phenomena studied here take now place upon variation of
this parameter: endemic states and full synchronization occur above
a certain critical value of the recombination rate. Moreover, for
the SIS epidemiological model with a given infectivity, we have
found that the limit of infinitely large recombination rate is
equivalent to the situation where the interaction pattern is static
but not disconnected. For chaotic maps, on the other hand, the fact
that time elapses by discrete steps imposes a limit to the
recombination rate: any interacting couple can at most be recombined
once per step. This establishes in turn an upper limit for the
Lyapunov coefficient of individual maps such that they can be
synchronized by recombination. If the maps are ``too chaotic'',
synchronization is not possible even at the maximal rate of one
recombination per couple per time step.

The fact that, instantaneously, each dynamical element of the two
ensembles considered here has just only one interaction partner,
makes the corresponding problems analytically tractable to a large
extent. In particular, we have obtained analytical approximations
for the critical values of the recombination rate which are in very
good agreement with numerical results. For SIS epidemics, the
critical recombination rate was found both from the integration over
the whole population of the epidemiological dynamics within each
couple, and from the dynamics of the number of couples in each
epidemiological state. For the chaotic maps, we have replaced this
latter approach by a kind of symbolic schematic representation of
the dynamics of couples. In this representation, which can also be
adapted to the SIS model and qualitatively reproduces the critical
behaviour of the two systems, maps (or epidemiological agents) are
though of as two--state elements. Their interactions induce
transitions between the two states following a small set of
intuitive rules. We conjecture that this kind of representation is
capturing the essential mechanisms that govern the relative
prevalence of unsynchronized/susceptible elements at one side of the
critical point,  and of synchronized/infectious elements at the
other.

Collective dynamics on monogamous interaction patterns have the
advantage of analytical tractability. It should however be borne in
mind that these patterns represent a kind of extreme case among
disconnected networks: they have the minimal number of links that
avoids isolated elements. Networks with less severe lack of
connectivity should impose lower limitations to the development of
collective self-organized behaviour. In these cases, therefore, we
expect that recombination, even at lower rates, will also be able to
compensate the lack of connectivity, triggering critical phenomena
such as those studied here.

\end{document}